\documentclass[pra,preprint,showpacs,showkeys,amsfonts]{revtex4}
\usepackage{graphicx}
\RequirePackage{times}
\RequirePackage{mathptm}
\begin{document}

\def\ttt{{\rm Tr} }
\def\diag{{\rm diag} }

\title{The dangerous misconceptions of Sir Karl Raimund Popper}
\author{Karl Svozil}
 \email{svozil@tuwien.ac.at}
\homepage{http://tph.tuwien.ac.at/~svozil}
\affiliation{Institut f\"ur Theoretische Physik, University of Technology Vienna,
Wiedner Hauptstra\ss e 8-10/136, A-1040 Vienna, Austria}

\begin{abstract}
Insofar as Sir Karl Raimund Popper's writings deal with political statements, they are evident;
yet insofar as they deal with scientific issues, they are misleading.
If applied to the concrete implementation of science,
such as distribution of research funds and (peer) review, they would seriously
impede progress.
\end{abstract}


\maketitle

There is little doubt that Sir Karl Raimund Popper is one of the most prominent figures
in the philosophy of science of the past century.
Many people, from politicians to scientists, pay lip service
to his theses, and numerous television shows, conferences and books have
been organized in his honour.
But while Sir Karl Raimund Popper has emphasized a
lot of things which are undoubtedly important,
his thoughts about the induction problem \cite{popper},
his criterion of demarcation between science and ``pseudo-science,''
at least if taken naively and at face value,
may be an impediment to scientific research programs, thereby resulting in a
waste of efforts and money.

In what follows,
I shall begin with a few brief remarks on his views on politics and his criticism of psychoanalysis.
The main subject of this article is a critical review of
Sir Karl Raimund Popper's ``solutions to the problems of induction''
and its consequences for the public understanding of science.
Concrete implementations of these ``solutions'' may
severely hamper the scientific progress by imposing too heavy and
detrimental criteria on science proper.
Finally, I shall try to discuss the question of
why Sir Karl Raimund Popper is so highly honoured in certain
circles; mainly in politics and in the natural sciences.

Most of what I say here has already been expressed many times by
philosophers of science such as, for example, Imre Lakatos
\cite{lakatosch} or Paul Feyerabend \cite{feyerabend,feyerabend1}.
But many of these critical issues are not mentioned in the almost
frenetic, uncritical appraisals and reviews of the work of Sir Karl Raimund Popper
which accompany the  100 anniversary of his birthday here in Vienna and elsewhere.
There are, however, some notable exceptions. Asher Peres has criticised
a Gedankenexperiment
devised by the late Sir Karl Raimund Popper to ``falsify'' the dominant Copenhagen interpretation
of quantum mechanics \cite{2002-peres}.
Marisa Dalla Chiara has given a critical account of Sir Karl Raimund Popper's
disapproval of quantum logic \cite{dalla-2002}, in particular the approach chosen by
Garrett Birkhoff and John von  Neumann \cite{birkhoff-36}.

Sir Karl Raimund Popper's thinking was deeply rooted in
the spiritual life of post-World War I Vienna in many ways.
He first felt attracted by some major schools of thought
which flourished in or even originated from this city,
but afterwards became one of their strongest opponents.
(Stated polemically, no school of thought should be eager to have Sir Karl Raimund Popper as a disciple.)

Having been a Marxist, he disguised Marxism as a metaphysical,
nonscientific ideology which brings misery to the masses.
Indeed, this fact must have appeared rather obvious, in particular
at the end of the World War II,
given the developments in the Bolshevik dominated Soviet block under
Stalin.
The same holds true for a criticism of Nazism and its various totalitarian
conservative offsprings, also in Austria before the German invasion.
Undemocratic ideologies had just rampaged through Europe and the Pacific.
(I personally think that, in view of the atrocities and inhumanities committed,
a lack of falsifiability is one of the less malign deficiencies of these kinds of totalitarianism.)

Back in his old Vienna days, Sir Karl Raimund Popper
also became fascinated by the psychoanalytic theories of Freud and Adler,
under whose aegis he engaged briefly in social work with children.
He later on denounced the psychoanalytic theory as nonscientific and useless,
mainly because it appeared to him to be difficult to falsify (or if
falsifiable, had been falsified).
This shock struck the psychoanalytic community, in particular its academic sections, hard.
I still remember joint sessions of the two main Vienna psychoanalytic societies
a couple of years ago,
which tried to cope with this criticism, which at that time had been mainly put
forward by Adolf Gr\"unbaum \cite{gruenbaum-psychoanalysis}.
It is not too unjustified to claim that Sir Karl Raimund Popper managed to academically discredit
psychoanalysis up to this date, at least what its influence in the
academic world is concerned.
In academic psychology, phenomenologically oriented and ``falsifyable''
theories such as behaviorism flourished;
 which were ``scientifically sound'' and reasonable according to
Sir Karl Raimund Popper's criterion of demarcation.
I have strong doubts that practical clinical psychology has benefited from this climate of thought.
On the contrary, my feeling is that this development has hampered or even stopped
in-depth, sustainable, treatments and cures of mental illnesses
of epidemic proportions such as depression and neurosis.

Sir Karl Raimund Popper's main contribution to the philosophy of science has been his
 ``solutions to the problems of induction.''
This was conceived against another Viennese invention: against the ``verification principle''
of the Vienna Circle
(which included Moritz Schlick, Rudolf Carnap, Otto Neurath, Viktor Kraft, Hans Hahn, and Herbert Feigl).
He turned the verification principle upside down and
introduced a ``falsification  principle'' as a demarcation criterion
between science proper or ``quasi-science,'' or, as he sarcastically used to called it, ``blablabla.''

While it is quite obvious that no operational verification can ever
prove
a scientific theory to be ``true,''  it is not totally clear why a necessary
criterion for any thought to be considered scientific should be its falsifiability.
Indeed, Imre Lakatos repeatedly pointed out \cite{lakatosch} that
such a demarcation criterion assumes that there are critical tests,
which are able to conclusively falsify a theory.
History tells us that many operational test of a theory could be
interpreted in
one way or the other---either as corroboration of a theory or as a falsification.

Moreover, a theory or research program {\it in statu nascendi} is almost always handicapped
by a badly developed formalism and most of the time lacks the experimental credentials
of its soon-to-become predecessors. A historical example which is often mentioned is the Copernican system,
putting the earth in rotation around the sun, which
was competing against the older Ptolemaic system.

Sir Karl Raimund Popper himself has conceded these and other problems with the demarcation criterion.
And he seemed to have constantly revised and adapted the demarcation
criterion to cope with the historic facts.

One may quite justifiable ask if some principles of scientific conduct
which claim to be able to differentiate and decide between what is reasonable science
and ``blablabla,'' itself should not also satisfy the very principles
it requires from proper science. In other words: is Sir Karl Raimund Popper's
theory of induction a scientifically sound theory by its own standards, or mere ``blablabla?''

While I shall let the Reader decide the case, I would only like to mention
my impression that Sir Karl Raimund Popper seemed to have applied the same kind of
``immunizing strategies''
which he blamed other ``quasi-scientific'' constructions
to  his own theories.
In short, the demarcation criterion may just be an ideology according to its own judgements, after all!

It is interesting to note that some of Sir Karl Raimund Popper's very few and
almost forgotten early papers in physics (e.g.,
Refs. \cite{popper-34,popper-50i,popper-50ii})
avoid concrete predictions and statements which could be falsified.
Should they therefore considered be as ``blablabla?''
I do not think so; they contain highly original speculations
about the consequences of G\"odel's incompleteness theorems for physics.

If the demarcation criterion is an ideology,
then the old Roman dictum ``{\it cui bono?}'' applies;
i.e., who are the losers and the winners?
In more practical terms, almost every decision has a financial component
associated with it. To put it pointedly:
if one omits the scientific content and meaning of a decision for the time being,
then a decision often boils down to money.
If a wrong or a misguided ideology is taken as a guideline for decision
making such as funding policies, then this boils down to a waste of
(mostly taxpayer's) money.
Hence,  Sir Karl Raimund Popper's claim to be able to draw a
demarcation line between proper scientific conduct and pseudo-science
may not only impede the growth of knowledge, but may also be
very costly.

However, as mentioned already, most of the practicing scientists,
although paying lip service to the principle of falsification,
do not really care about it in their everyday operations.
The following anecdote may serve as an example.
A physics department head once boldly stated,
`if a student comes to me with a new idea, I first ask the student
if the idea is falsifiable, at least in principle.'
If not, he concluded, this idea had to be dismissed immediately
(this seemed almost a relieve to him).
Fortunately, the researcher did not implemented the strategy
just mentioned in his home institution.
Otherwise, more than one half of the ongoing research there would have to be abandoned.

Indeed, if the demarcation criterion would be implemented into theoretical physics,
many venerable research journals would have to stop publishing.
And yet, ``peer review,'' with its assumed benevolent
censorship, or the formation of interests clusters and pressure groups
(``gangs'') effectively might do more harm and might be responsible for
a bigger waste of
(mostly taxpayer's) money than the demarcation criterion.
This could be
the topic of another
article.
Here, I would only like to mention that I believe it is not totally unreasonable to try to utilize
other forms of evaluation of research than just ``peer review,''
in particular an implementation of the grand jury system as it is
already practiced
in the courtrooms. This has been proposed by Paul Feyerabend, yet
another philosopher of science originating from Vienna. But I would
also like to propose
to distribute a certain amount of money to research programs by a random
selection, such as a lottery throwing dices; with a post mortem
evaluation of all three funding groups
(peer review/jury selected/randomly selected; maybe distributed by a
70:20:10 ratio).

My impression is that any kind of generally applicable
demarcation criterion between
proper science and ``pseudo-science'' eventually will turn out to be a
red herring.
In the same category are attempts to implement Occams razor as
a criterion (c.f. the criticism by Daniel Greenberger
\cite{greenberger-occamsrazor}),
or the radical operationalism of the Percy Bridgman \cite{bridgman36}.

The question remains open
why the thoughts of Sir Karl Raimund Popper have been so highly appreciated.
In my opinion, as far as politics is concerned, this esteem comes from the fact that he fitted nicely
into the scheme of liberal democracy at a time when western policy
makers demanded someone who would firmly stand against communism and for western liberal democratic values,
 while at the same time
had not the faintest touch of Nazism (such as, for instance, Heidegger and Adorno).
This may explain some of the fame of Sir Karl Raimund Popper in
political circles.

But this cannot explain the high reputation among scientists, although
he himself claimed not to have been taken too seriously by his fellow
colleagues in philosophy departments, contrary to researchers in
physics, medicine, biology and other natural sciences.---Maybe the former were too
sophisticated to agree and also suspicious what might remain of their own
research after Sir Karl Raimund Popper's demarcation criterion
had been applied to their subjects.

Maybe one should view much of Sir Karl Raimund Popper's
Sir Karl Raimund Popper's ``solutions to the problems of induction''
as having a high marketing
value: it may be wise to agree officially to these almost undeniable,
self-evident statements, while at the same time pursuing many other
different goals, both politically and scientifically.
Yet, I have to confess not having found any convincing answer so far.
The high appreciation Ludwig Wittgenstein (yet another
Viennese philosopher) enjoyed in Cambridge is comparable, although this
might be explained by the ``Nostradamus-like'' style of his writings and
expressions, which leaves open the doors for many, even
contradicting, interpretations, and which sometimes appears to be
extremely attractive to philosophers.
If one considers Niels Bohr's philosophical emanations on the
interpretation of quantum mechanics (and his discussions with Einstein
and others on these issues), one is almost inclined to enlarge this
group to physicists as well.

Undoubtedly, many of  Sir Karl Raimund Popper's thoughts are insightful
and prudent; for instance when it comes to a radical revision
of the university system,
suggesting that the teaching of students (in New Zealand and elsewhere)
should be primarily based on the research spirit rather than on
grading ever larger numbers of
students \cite{aefppp}---an old Humboldian approach
which tends to become forgotten and has to be emphasized over and over again.
And I would have been willing to write
a more forgiving review (or none at all) of his theses
if he would have been more forgiving to his rivals and to the theories he
considered to be ``pseudo-science'' or simply ``blablabla.''
Unfortunately, this was not the case.
And hence, critical remarks seem to be in order.

Let me conclude this very brief and necessarily incomplete appraisal
of the works of Sir Karl Raimund Popper with his statement,
``all theories are hypotheses; all may be overthrown,'' which serves as
an almost tautological motto to an exhibition advertised by the University
of Vienna in connection with its recent Karl Popper 2002
Centenary Congress.
Or should one better recall Ludwig Wittgenstein's dictum, ``Wovon man
nicht sprechen kann, dar\"uber mu\ss~ man
schweigen'' (``What we cannot speak about we must
pass over in silence'')?

\subsection*{Acknowledgements}
The author is grateful to Garry Tee for bringing to his attention
reference \cite{aefppp}.


\end{document}